\documentclass[11pt]{article}
\usepackage[utf8]{inputenc}
\usepackage[margin=2.5cm]{geometry}
\usepackage{graphicx}
\usepackage{multirow}
\usepackage{comment}
\usepackage{url}
\usepackage[dvipsnames]{xcolor}
\usepackage[normalem]{ulem}
\usepackage[version=4]{mhchem}
\usepackage{amssymb}
\usepackage{amsfonts}
\usepackage{amsmath}
\usepackage{authblk}
\usepackage{mathtools}

\DeclareMathOperator{\var}{var}
\DeclareMathOperator{\cov}{cov}
\newcommand{\expect}{\mathbb{E}}
\newcommand{\pdf}[2][]{\frac{\partial{#1}}{\partial{#2}}}

\DeclarePairedDelimiterX{\infdivx}[2]{[}{]}{%
  #1\;\delimsize\|\;#2%
}
\newcommand{\infdiv}{D\infdivx}

\title{Determining the maximum information gain and optimising experimental design in neutron reflectometry using the Fisher information}
\author[1]{James H. Durant}
\author[2]{Lucas Wilkins}
\author[3]{Keith Butler}
\author[1]{Joshaniel F. K. Cooper}
\affil[1]{\small{ISIS Neutron and Muon source, Rutherford Appleton Laboratory, Harwell Campus, OX11 0QX}}
\affil[2]{\small{Department of Zoology, University of Oxford, Mansfield Road, Oxford, OX1 3SZ}}
\affil[3]{\small{SciML, Scientific Computing Division, Rutherford Appleton Laboratory, Harwell Campus, OX11 0QX}}
\date{}

\begin{document}
\maketitle
\begin{abstract}
An approach based on the Fisher information (FI) is developed to quantify the maximum information gain and optimal experimental design in neutron reflectometry experiments. In these experiments, the FI can be analytically calculated and used to provide sub-second predictions of parameter uncertainties. This approach can be used to influence real-time decisions about measurement angle, measurement time, contrast choice and other experimental conditions based on parameters of interest. The FI provides a lower bound on parameter estimation uncertainties and these are shown to decrease with the square root of measurement time, providing useful information for the planning and scheduling of experimental work. As the FI is computationally inexpensive to calculate, it can be computed repeatedly during the course of an experiment, saving costly beam time by signalling that sufficient data has been obtained; or saving experimental datasets by signalling that an experiment needs to continue. The approach's predictions are validated through the introduction of an experiment simulation framework that incorporates instrument-specific incident flux profiles, and through the investigation of measuring the structural properties of a phospholipid bilayer.
\end{abstract}

\section{Introduction}
The Fisher information (FI) \cite{Fisher1925} has been applied across many fields, from information theory and communications \cite{Wang2010, Barnes2019} to quantum mechanics \cite{Barndorff-Nielsen2000, Petz2002}, quantitative finance \cite{Taylor2019} and volcanology \cite{Telesca2009}. The FI provides a way of measuring the amount of information an observable variable carries about an unknown parameter of a distribution that models the observable. For certain situations it is possible to analytically calculate the FI, giving a measure of parameter uncertainty, as well as inter-parameter covariances, from which correlations can be derived. Neutron reflectometry allows one to model a measured reflectivity curve in order to determine the properties of the thin film layer structure that produced the curve. Most reflectometry analyses use sampling methods to extract parameter uncertainties, though this is expensive and cannot be performed in real time with current software \cite{Nelson2019, Refl1D, RasCAL}. In this work, we describe an application of the FI to neutron reflectometry in enabling real-time estimation of parameter uncertainties, as well as a projection of these with time. We compare the results to established sampling methods and demonstrate the FI's use for experimental design, and for potentially enabling early stopping of experiments based on counting statistics.

In reflectivity, a thin film is described by a thickness, a scattering length density (SLD) which is the product of the neutron scattering length and the film density, and an interfacial roughness; thin film heterostructures are composed of multiple thin films on top of each other. In the analysis of reflectometry data, we are presented with data points of reflectivity, $r$, as a function of momentum transfer, $Q$, and wish to infer the SLD profile from the top surface to the substrate. For a single interface, i.e. a semi-infinite substrate, the neutron reflectivity decays $\sim Q^{-4}$, also called the Fresnel reflectivity. A single layer on a substrate is analytically solvable \cite{Sivia2013}, however, for more layers, multiple reflections are possible, and the inversion of the curve to a SLD profile is non-trivial. In fact, the loss of phase information upon reflection makes inversion of the SLD profile from the reflectivity profile an inverse problem \cite{Majkrzak1995} and approximations are required.

Typically, reflectometry analysis is model-dependent where a model is defined using series of contiguous layers and the model reflectivity is calculated using the Abel{\`{e}}s matrix formalism for stratified media \cite{Abeles1948} or Parratt recursive method \cite{Parratt1954}. However, the solution to this analysis is not necessarily unique and often requires \textit{a priori} knowledge such as details of a system or the underlying science. Such prior knowledge helps to limit the dimensionality of the optimisation space by reducing the number of structures that agree with the experimental data within some tolerance. Methods have been devised to estimate interface properties, using this prior knowledge, that describe the data while adhering to a given set of constraints. Such methods include optimisers applying gradient projection \cite{Byrd1995}, annealing processes \cite{Xiang1997}, and evolutionary algorithms \cite{Storn1997}.

Another approach to optimisation is the use of sampling methods of which there are two discussed in this work, Metropolis-Hastings Markov Chain Monte Carlo (MCMC) \cite{Metropolis1953, Hastings1970} and nested sampling \cite{Skilling2004, Skilling2006}, both of which are Bayesian and sample the parameter posterior distribution. Due to the typically high dimensionality of the parameter space in reflectometry, Bayesian sampling methods tend to be computationally expensive and impractical for obtaining results, such as parameter estimates and covariances, in real-time. We use \texttt{refnx} MCMC sampling \cite{Nelson2019} and \texttt{dynesty} nested sampling \cite{Speagle2019} to sample our data and to compare the results to those derived from the FI; \texttt{refnx} uses the \texttt{emcee} package \cite{Foreman-Mackey2012} to provide an implementation of an invariant MCMC ensemble sampler \cite{Goodman2010}.

Much work has been undertaken on quantifying the information content of a reflectivity dataset, with most applying Bayesian statistics, where probability represents a degree-of-belief or plausibility based on the evidence at hand \cite{Sivia2006}. One such approach, looked at experimental optimisation by determining the information gain from a given experiment using the entropies of the posterior and prior probability density functions \cite{Treece2019}. Similarly, work has been done on quantifying the information gain from scattering experiments as a function of the scattering length density of molecular components \cite{Heinrich2020}. Many other Bayesian information-based approaches have been applied to reflectometry including the use of Bayesian evidence to determine the set of free parameters that maximise model information density \cite{McCluskey2020}, and using maximum entropy to reconstruct a SLD profile from a reflectivity curve \cite{Sivia1991}.

Similarly to previous works, we propose a methodology for quantifying the information content of a reflectivity dataset for use in determining maximum information gain and experimental design optimisation. However, we attempt to solve a slightly different problem to previous work; the calculations that are made using our framework are different to those of Bayesian techniques. Although the goal of these estimation procedures remains the same, we derive the maximum information that the dataset contains, given the current data point error bars, not the information content that can be readily extracted; for example, by sampling the posterior distribution.

For our application, the uncertainties on our reflectivity points are defined as the square root of the number of neutron counts and these counts are governed by Poisson statistics. Under these assumptions, we can analytically calculate the FI and apply the Cram\'{e}r-Rao bound \cite{Cramer1946, Rao1994}. The bound states that the inverse of the FI provides a strict lower bound on the variance of an observable variable and, as consequence, the FI provides us with a strict upper bound on the amount of information extractable from the observable. In practice, using this analytical derivation, we can achieve sub-second calculations of parameter uncertainties.

To evaluate the FI approach, we developed an experiment simulation framework based on the underlying assumptions of Poisson statistics for neutron counts. Since the FI is calculated using neutron counts, such a framework is necessary in calculating the FI in the general case where any model is given. Further, this framework allows us to calculate the information content of any experimental conditions without costly beamtime to acquire the same data. The simulation framework is general in that it can simulate any beamline, given the incident flux profile of the instrument in question, and shown to be accurate without requiring computationally expensive Monte Carlo methods.

\section{Methods}
\subsection{The Fisher Information Matrix}
The FI is a fundamental quantity in statistical theory and quantifies parametric changes in probability distributions in terms of information. It is related to various kinds of information ``distance'', most notably the Kullback-Leibler (KL) divergence: a building block for many familiar information theoretic measures \cite{Kullback1968}. The KL divergence is the standard way of measuring a difference between distributions in terms of information and, when applied to parametric distributions, provides the ``distance'' between parameter values. As a rule of thumb, we can understand a one sigma (one standard deviation) difference in parameters as a KL divergence of one \textit{nat} (natural unit of information). To a first approximation, this distance is calculated from the FI.

More formally, the FI matrix, $\mathbf{g}$, is the first non-zero term in the series expansion of the KL divergence from one (vector) parameter $\xi$ to $\xi + \Delta\xi$ (written here as $\infdiv{\xi}{\xi + \Delta\xi}$):
$$
\infdiv{\xi}{\xi + \Delta\xi} =  \frac{1}{2}\Delta\mathbf{\xi} \mathbf{g} \Delta\mathbf{\xi}^T + O(|\Delta\xi|^3)
$$
Thus, we can think of the FI matrix as a way of scaling the parameters so that, for sufficiently small changes in parameters, the square Euclidean distance is the informational change. The FI is therefore relative to the specified parameters, and measured in \textit{nats} per parameter unit squared; it is local and not dimensionless.

The one \textit{nat} per sigma relationship is exact for many widely used distributions, such as for a multivariate normal with constant covariance. In this case, the FI is the inverse of the associated covariance matrix. Correspondingly, a practical way this is used is to set an information threshold, e.g. one \textit{nat} (one sigma), and find by how much parameters must change to reach this threshold, thereby specifying an acceptance and/or confidence region, as is discussed in section \ref{FI_application} and the SI.

\subsection{Derivation}
To derive the equations for information content quantification using the FI, we must first provide a structure for given reflectivity data of $N$ points (equivalently, histogram bins). This structure consists of contiguous layers representing a physical sample with each layer being defined by its thickness, SLD and interfacia roughness. A model is then described by this structure and given measurement background noise, experimental scale factor, and instrument resolution; we need only vary the $M$ unknown parameters of this model. Such a model describes the reflectance at a given neutron momentum transfer, for example, using the Abel{\`{e}}s matrix formalism implemented in \texttt{refnx}. Please refer to the supplementary information (SI) for the full derivation but, in summary, the FI matrix, $\mathbf{g}^\xi$, for the $M$ parameters, $\xi$, of a model of $N$ reflectivity points, is given by
\begin{equation}
    \label{eqn:FI}
    \mathbf{g}^\xi = \mathbf{J^T}\mathbf{MJ}
\end{equation}
where $\mathbf{J}$ is the Jacobian of the reflectances, $r_i$, with respect to the parameters, $\xi$, and $\mathbf{M}$ is a diagonal matrix of incident counts, $s_i$ divided by model reflectances $r_i$. In the FI matrix, the FI for an individual parameter, $\xi_i$, corresponds to the diagonal element $\mathbf{g}^\xi_{i,i}$.

\begin{minipage}{0.5\linewidth}
    \begin{equation}
        \label{eqn:J}
        \mathbf{J} = \begin{bmatrix}
        \pdf[r_1]{\xi_1} & \pdf[r_1]{\xi_2} & \cdots & \pdf[r_1]{\xi_M}  \\
        \pdf[r_2]{\xi_1} & \pdf[r_2]{\xi_2} & \cdots & \pdf[r_2]{\xi_M}  \\
        \vdots           & \vdots           & \ddots & \vdots  \\
        \pdf[r_N]{\xi_1} &\pdf[r_N]{\xi_2}  & \cdots & \pdf[r_N]{\xi_M}  \\
        \end{bmatrix}
    \end{equation}
\end{minipage}
\begin{minipage}{0.5\linewidth}
    \begin{equation}
        \label{eqn:M}
        \mathbf{M} = \begin{bmatrix}
        s_1 / r_1 & 0         &   \cdots  & 0           \\
        0         & s_2 / r_2 &   \cdots  & 0           \\
        \vdots    & \vdots    &   \ddots  & \vdots      \\
        0         &0          &   \cdots  & s_{N} / r_N \\
        \end{bmatrix}
    \end{equation}
\end{minipage}

Using equation \ref{eqn:FI}, we can calculate the FI for a model describing a single dataset. However, more complicated experiments often involve multiple datasets, such as measuring multiple experimental contrasts. For such cases, $n$ models, potentially containing inter-model parameter constraints, are required as input. The calculation is the same as in equations \ref{eqn:FI}, \ref{eqn:J} and \ref{eqn:M} except the parameters, $\xi$, are the union of all of the (potentially shared) parameters of the $n$ models. Additionally, the Jacobian, $\mathbf{J}$, and matrix, $\mathbf{M}$, are calculated over the concatenated incident counts and model reflectances.

\subsection{Experiment Simulation}
To simulate an experiment, we require both a model and knowledge of the flux of incident neutrons as a function of wavelength. In our case, this was taken on the OFFSPEC reflectometer \cite{Dalgliesh2011}. We can multiply this incident neutron flux by a constant in order to change the experimental counting time and to give us the number of incident neutrons for a simulated experiment; this approach was developed from the ideas presented by \cite{Mironov2021}. To account for different measurement angles, we multiply this incident flux by a factor to compensate for different collimating slit openings. Since both of the slits that are used to define the beam footprint scale linearly with the angle, we scale the intensity as the square of the angle.

We calculate the momentum transfer, $Q$, for each wavelength, $\lambda$, in the file using the measurement angle, $\theta$, and equation
$$Q = \frac{4\pi \sin\theta}{\lambda}$$
By default, these $Q$ values are assigned to geometrically spaced bins, with the number of bins being set to the desired number of points for the simulated dataset. Following this, we calculate each bin's centre, $Q_i$. Alternatively, a given set of $Q$ bin centres can be used. The model reflectivity for each bin, $r_i$, is calculated using \texttt{refnx} and additive instrument background noise, $\alpha$, is added (optionally, this can be accounted for in the model reflectivity calculation). Next, this reflectivity is multiplied by the bin’s incident flux, $\mu_i$, to obtain the reflected flux and then multiplied by the simulated measurement time, $\tau$, to get the reflected counts for the bin. We use the reflected counts as the mean rate parameter of a Poisson distribution, from which, we obtain a random value giving us an appropriately randomised number of reflected counts, $N_i$.
$$N_i \sim \operatorname{Poisson}\big[ (r_i + \alpha)\mu_i\tau \big]$$
The bin's uncertainty in count-space is then the square root of this value, $\sqrt{N_i}$. To obtain the reflectivity and associated uncertainty, we simply divide the ``noisy'' counts and uncertainty by the number of incident neutrons for the bin, $s_i$ (I.e. the product of the bin's incident flux and measurement time, $\mu_i \tau$).

The previously described process will generate a single reflectivity dataset for a given model. For more complicated experiments involving multiple datasets, $n$ models are required as input, and the above process repeated for each model, yielding $n$ simulated datasets.

\subsection{Applying the Fisher Information} \label{FI_application}
The FI can be applied to both experimentally measured and simulated data. As the FI measures the instrumental uncertainty relative to changes in model parameters, a parameterised model must be provided and parameter values input. These parameter values can be obtained as estimates based on data, or specified manually if the task is simply the verification of a particular structure.

We use \texttt{refnx} to define a model and to load the model's associated measured or simulated reflectivity data. The parameters of the model are then optimised using a fitting algorithm of choice, in our case differential evolution \cite{Storn1997}, according to the $\chi^2$ distance or equivalently, the negative log-likelihood. The likelihood provides a measure of difference between a given dataset and model and, for this work, is defined as its implementation in \texttt{refnx}
$$
-\ln{\mathcal{L}} = \frac{1}{2} \sum_{i=1}^{N} \Bigg(
                                                    \bigg( \frac{r_i - r_{i_m}}{\delta r_i} \bigg)^2 +
                                                    \ln{\big[ 2\pi(\delta r_i)^2 \big]}
                                               \Bigg)
$$
where, $N$ is the number of measured data points, $r_i$ is the experimental reflectivity at the $i$th $Q$ point, $\delta r_i$ is the uncertainty in the experimental reflectivity at the $i$th $Q$ point and $r_{i_m}$ is the model reflectivity at the $i$th $Q$ point calculated using the Abel{\`{e}}s matrix formalism.

From here on, we no longer need the data, since the model and Poisson statistics describe the data sufficiently. Next, we calculate the Jacobian, $\mathbf{J}$, whose entries are the gradient of the model reflectivity, $r_i$, with respect to each of the model parameters, $\xi_j$. We estimate this by using a finite difference approximation based on the reflectance for parameter values 0.5\% either side of the input value. Using this and the diagonal matrix, $\mathbf{M}$, of incident counts divided by model reflectances, we can calculate the FI matrix using equation \ref{eqn:FI}. Since, the calculation of this matrix is relatively simple, implementation for use in other fitting software \cite{Refl1D, RasCAL, Bjorck2007} should be straightforward.

The FI matrix contains all of the information about parameter variances and covariances but these values require extraction. The variance of a single parameter is simply given by the inverse of the FI and so its uncertainty is given by the square root. In the general case of multiple parameters, the uncertainty, $\epsilon_i$, for a parameter, $\xi_i$, is obtained from the square root of the inverse of the diagonal elements of the FI matrix
\begin{equation}
    \label{eqn:FI_uncertainty}
    \epsilon_i = \sqrt{1 / \mathbf{g}^\xi_{i,i}}
\end{equation}
Finally, to extract the covariance between any two parameters, we can calculate a confidence ellipse of given size $k$ standard deviations (see SI for details).

\subsection{Application to Soft Matter} \label{softmatter_methods}
To illustrate the utility of the FI, we applied our framework to an experiment measuring a common model system for structural biology: a 1,2-dimyristoyl-\textit{sn}-glycero-3-phosphocholine (DMPC) bilayer deposited onto a silicon surface. The lipids were measured against two water contrasts, \ce{H2O} and \ce{D2O}. The data was taken using the CRISP neutron reflectometer \cite{Penfold1987} as part of the ISIS neutron training course and simultaneously fit using \texttt{RasCAL} \cite{RasCAL}. This fitting was constrained against measured data for a bare \ce{Si}/\ce{D2O} interface including a native \ce{SiO2} layer.

Our model for the bilayer was defined by two lipid leaflets with fixed surface coverage. The model was fitted by area per molecule rather than volume fractions to avoid ambiguity arising from differing total molar quantities of headgroup and tailgroup components. The model also accounted for the headgroups and tailgroups containing water through defects across their surfaces and also the water bound to the hydrophilic headgroups. After fitting the experimental data, we reparamaterised the bilayer model as a function of contrast SLD and, using this new model, were able to simulate the DMPC bilayer experiment on the OFFSPEC reflectometer (using our instrument flux profile) with arbitrary contrast SLD. We then investigated the change in the FI, for each model parameter, with contrast SLD.

For our parameterisation, we have assumed the molecular volumes of the headgroups and tailgroups are known and constant, and that any changes in molecule surface area are inversely proportional to the headgroup and tailgroup thicknesses. Structural biology is a large field of research with varying values for these molecular volumes used. Further, these molecular volumes may vary with measurement conditions and may not necessarily be constant in practice \cite{Campbell2018}. However, as to not overcomplicate our model, we have fixed them. The full details of the bilayer model parameterisation and fitting can be found in the SI.

\section{Results and Discussion}
\subsection{Measured vs. Simulated Data}
To demonstrate the robustness of our experiment simulation, we compare a dataset measured using the OFFSPEC neutron reflectometer to its simulated counterpart. The data was experimentally measured using angles $0.3^{\circ}$, $0.4^{\circ}$, $0.5^{\circ}$, $0.6^{\circ}$, $0.7^{\circ}$, $2.0^{\circ}$ and $3.0^{\circ}$ with measurement times 7.5, 7.5, 7.5, 15, 15, 60, and 120 minutes respectively. The data from these angles was stitched together to produce a single data file. To obtain a ``ground truth'' model for simulation, we fitted this stitched data using \texttt{refnx} to get table \ref{tab:measured_fitted}. The background, experimental scale factor and resolution used for fitting were $8\times 10^{-7}$, 0.783, and 2.5\% $dQ/Q$ respectively.

\begin{figure}
    \includegraphics[width=0.9\textwidth]{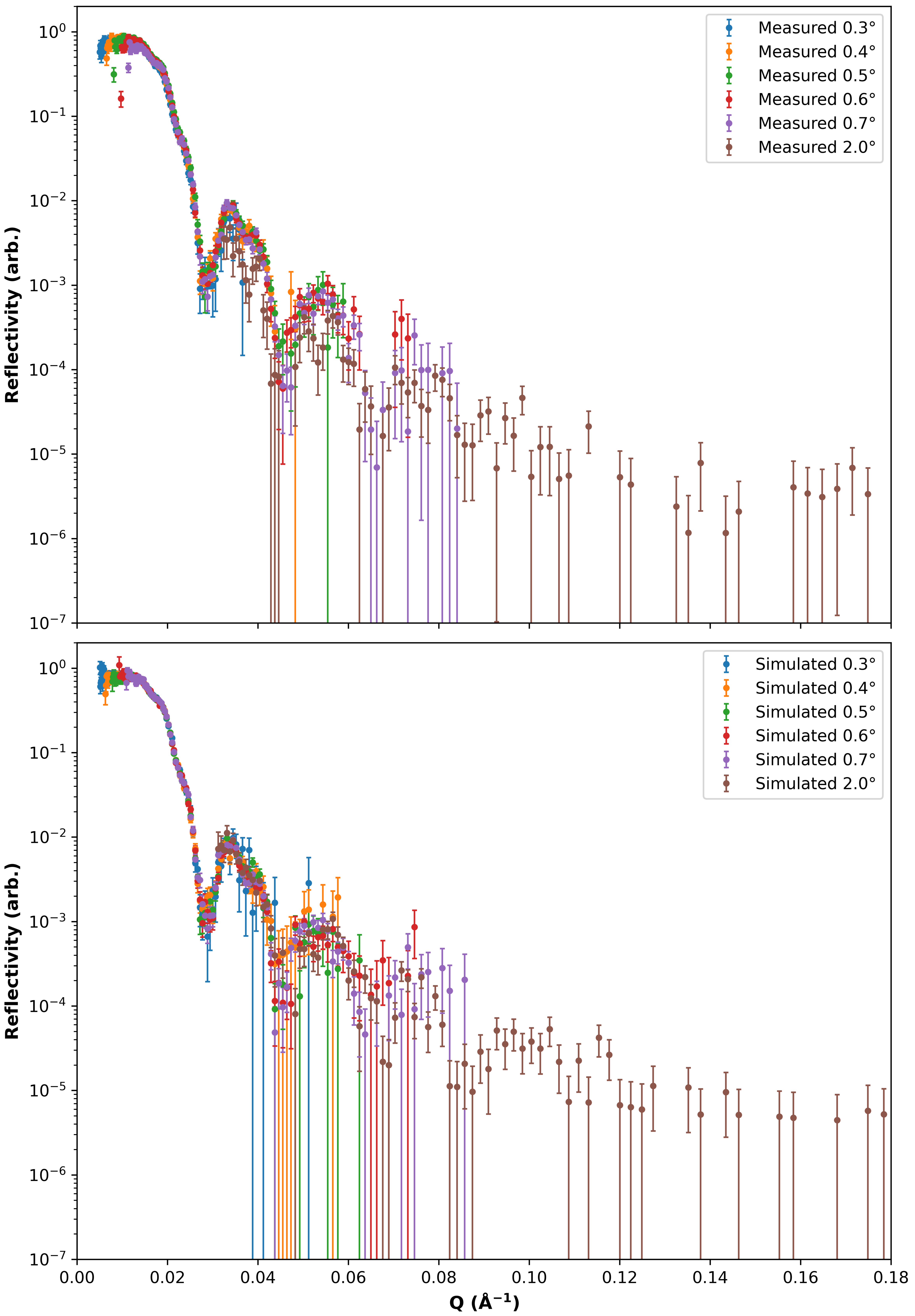}
    \caption{Experimentally measured reflectivity (top) and simulated reflectivity (bottom) vs. momentum transfer, $Q$, for each measurement angle of the table \ref{tab:measured_fitted} sample.}
    \label{fig:measured_data}
\end{figure}

To facilitate a measurable difference in the noise characteristics of the experimentally measured data and the data generated by our simulation framework, we took 1.5 minute time-slices from the measured data associated with each individual angle. For each measurement angle, we used the same angle, counting time and number of points for the simulation. As can be visually seen in figure \ref{fig:measured_data}, the noise characteristics of the time-sliced measured data, and simulated data are very similar. Statistically, comparing the datasets we find using the Hotelling's $t^2$ test $p=0.874$ and $t^2 = 0.159$, Anscombe Transformed \cite{Hotelling1931, Anscombe1948}, implying no significant differences between the measured and simulated data.

\begin{table}[!h]
    \centering
    \begin{tabular}{|l|l|l|l|}
    \hline
                       & SLD ($10^{-6} \text{\AA}^{-2}$) & Thickness ($\text{\AA}$) & Roughness ($\text{\AA}$) \\ \hline
    Layer 1 (\ce{Si})  & 1.795                           & 790.7                    & 24.5                     \\ \hline
    Layer 2 (\ce{Cu})  & 6.385                           & 297.9                    & 3.50                     \\ \hline
    Substrate (Quartz) & 3.354                           & N/A                      & 12.9                     \\ \hline
    \end{tabular}
    \caption{Fitted SLD, thickness and roughness values for each layer of the model corresponding to the measured dataset.}
    \label{tab:measured_fitted}
\end{table}

\subsection{Benchmarking} \label{benchmarking}
As mentioned previously, the FI approach has notable performance upsides. To demonstrate this, we compared the time to obtain parameter uncertainties using the approach and using established Bayesian methods, given a correct and fitted model. Fitting times have been excluded from these results since the time to fit would dominate the computation time of the FI approach, and thus would provide little insight into the computational advantage of the FI calculation. Additionally, since fitting is typically required for MCMC sampling but not for nested sampling, a fair comparison between all methods becomes difficult. We therefore focus on the time taken for each method to complete given optimal starting values.

The benchmark was run on a CPU with no methods having multiprocessing explicitly enabled. MCMC sampling was run with a 400 step burn-in period followed by a 30 step sample with each sample being separated by 100 iterations. Nested sampling was run using the default \texttt{dynesty} stopping criteria which is optimised for evidence estimation \cite{Speagle2019}. Uniform priors were used with a 25\% bound above and below the ground truth for each parameter. Following this, we ran our FI approach on the same samples. Table \ref{tab:benchmark} compares the mean processing times of 10 samples for each number of layers and, as can be clearly seen, the FI approach is significantly faster. It should be noted that our implementation is not particularly optimised and we believe further performance gains could be obtained if they were required.

For each number of layers in the interval $[1,6]$, we randomly generated 10 samples and varied the SLD, thickness and interfacial roughness of each layer in each sample. Each sample used a silicon substrate of SLD $2.047 \times 10^{-6} \text{\AA}^{-2}$ and the random SLD, thickness and roughness of each layer were sampled from uniform distributions of intervals $[-1,10] \times 10^{-6} \text{\AA}^{-2}$, $[20,1000]\ \text{\AA}$ and $[2,8]\ \text{\AA}$ respectively. Using our experiment simulation, we synthesised data for each of these samples and ran both MCMC and nested sampling to obtain parameter uncertainties. Each experiment simulation consisted of 140 points obtained from two angles, $0.7^{\circ}$ and $2.0^{\circ}$, using simulated measurement times of 7.5 and 30 minutes  respectively. Background noise of $10^{-6}$, instrument resolution of 2\% $dQ/Q$ and experimental scale factor of 1.0 were used.

\begin{table}[!h]
    \centering
    \begin{tabular}{|c|c|c|c|c|c|c|c|}
    \hline
    \multirow{3}{*}{No. Layers} & \multirow{3}{*}{No. Parameters} & \multicolumn{6}{c|}{Calculation Time ($s$)}                                                                     \\
                                &                                 & \multicolumn{2}{c|}{MCMC Sampling} & \multicolumn{2}{c|}{Nested Sampling} & \multicolumn{2}{c|}{FI Approach} \\
                                &                                 & Mean              & SD             & Mean              & SD               & Mean            & SD              \\ \hline
    1                           & 3                               & 197.829           & 3.344          & 53.310            & 8.947            & 0.015           & 0.005           \\ \hline
    2                           & 6                               & 229.641           & 5.032          & 155.480           & 35.920           & 0.024           & 0.004           \\ \hline
    3                           & 9                               & 262.568           & 5.334          & 363.318           & 120.075          & 0.036           & 0.004           \\ \hline
    4                           & 12                              & 292.382           & 3.244          & 1968.574          & 124.743          & 0.047           & 0.004           \\ \hline
    5                           & 15                              & 330.579           & 9.531          & 2967.707          & 561.529          & 0.060           & 0.004           \\ \hline
    6                           & 18                              & 372.116           & 5.667          & 3862.186          & 700.430          & 0.076           & 0.005           \\ \hline
    \end{tabular}
    \caption{Calculation time of parameter uncertainties, in seconds, for MCMC sampling, nested sampling and the FI approach. For each number of layers, 10 samples were randomly generated using that number of layers with the mean and standard deviation of the calculation time recorded for each approach.}
    \label{tab:benchmark}
\end{table}

\subsection{Corner Plots and Confidence Ellipses}
In \texttt{refnx} and \texttt{dynesty}, the results of MCMC and nested sampling respectively can provide a corner plot which is ``an illustrative representation of different projections of samples in high dimensional spaces'' \cite{Foreman-Mackey2016}. These Bayesian sampling methods sample the parameter posterior distribution, allowing contours to be drawn through samples that are equally probable. The FI, however, is developed from a frequentist view and the confidence ellipses bound regions where we have at least a $k\sigma$ confidence of the value. Despite these fundamental differences, the sampling corner plots do still often agree very closely with the FI confidence ellipses.

\begin{figure}
    \includegraphics[width=\textwidth]{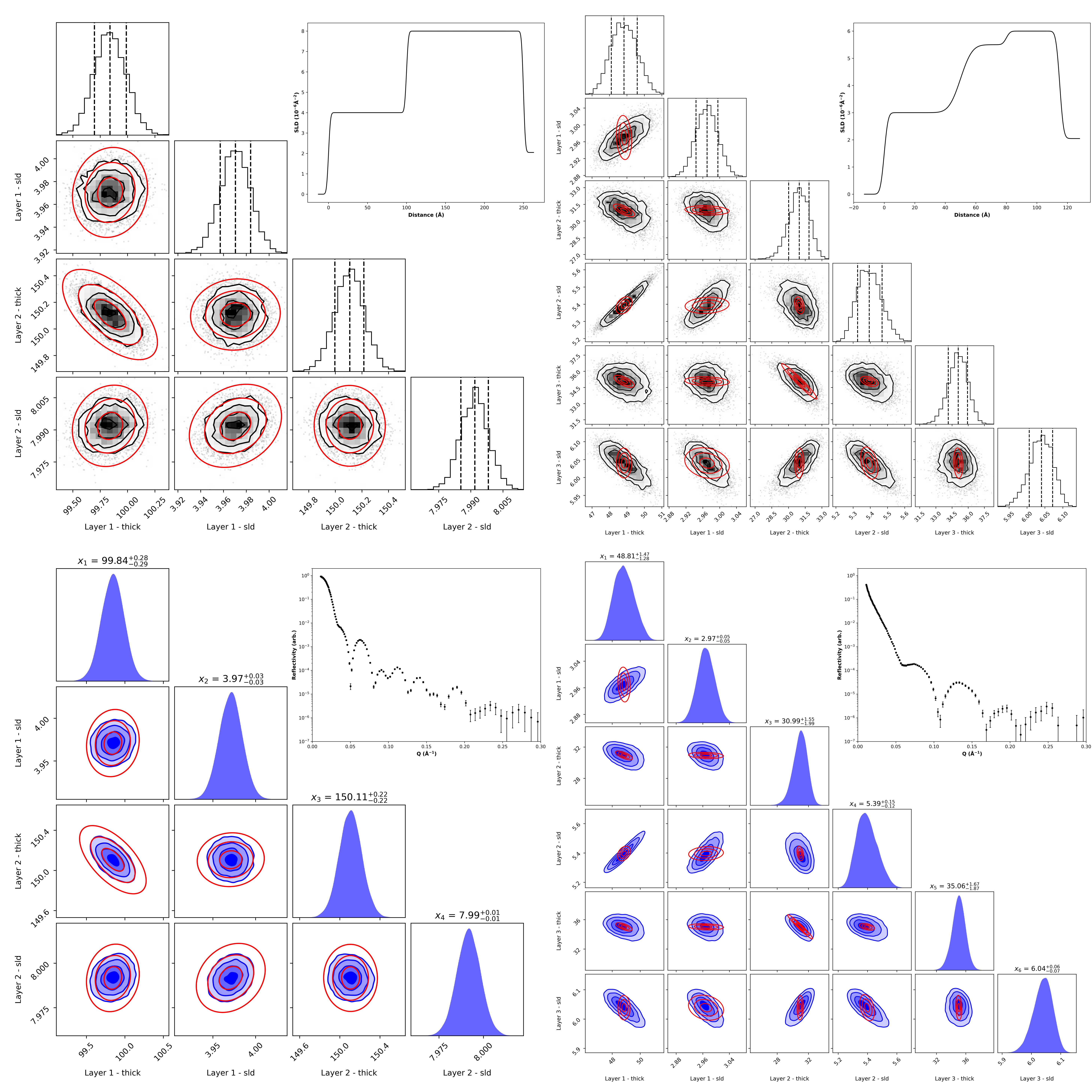}
    \caption{The Fisher information confidence ellipses for $k=1,2,3$ (red) overlaid on the corner plots of MCMC (black) and nested sampling (blue) for the mostly uncorrelated parameter sample (left) and correlated parameter sample (right). Inset are the SLD profiles (top) and rebinned simulated reflectivity curves (bottom) of the two samples.}
    \label{fig:confidence_ellipses}
\end{figure}

For samples with mostly uncorrelated parameters, we found that corner plots show strong agreement with confidence ellipses. However, when more parameter correlation is present, sampling uncertainties are much larger and we reach a point at which the FI still represents the maximum obtainable information, but this seemingly cannot not be extracted from the experimental data. As consequence, the confidence ellipses do not match the corner plots as closely. This discrepancy is shown in figure \ref{fig:confidence_ellipses} which compares two samples: a simple sample with mostly uncorrelated parameters and a more complicated sample with more parameter correlation due to similar layer SLDs. The datasets of figure \ref{fig:confidence_ellipses} were both simulated with the same run condition as detailed in section \ref{benchmarking}.

One potential source of deviation between corner plots and confidence ellipses may come from our fitting algorithm of choice. For our application of the FI, our estimator is a fitting algorithm and, so far, we have assumed that this estimator is unbiased. Thus, the Cram\'{e}r-Rao bound implies that the inverse of the FI is a lower bound on the variance of this estimator. However, in practice, we found that our fitting algorithm of choice, differential evolution in \texttt{refnx}, may exhibit bias in some cases. To measure this bias we simulated 1000 experiments, using the same simulation conditions previously used, for a number of different samples of varying complexity and calculated the difference between the ``ground truth'' and mean fitted parameter values. Table \ref{tab:biases} shows the fitting biases in the parameters of the figure \ref{fig:confidence_ellipses} samples. As can be evidently seen, the fitting bias is greater in the sample with larger inter-parameter correlations, particularly in the layer thicknesses; these biases are model dependent and potentially fitting package dependent.

The Cram\'{e}r-Rao bound may be modified for a biased estimator. However, for a real measurement, there is no way to be able to tell if such a bias exists. As such, we leave our approach with the stricter limit (since any bias always increases the variance), and remind ourselves that the maximum possible information contained in the data is not always going to be the maximum extractable.

\begin{table}[!h]
    \centering
    \begin{tabular}{|c|c|c|c|c|c|c|}
    \hline
    \multirow{3}{*}{Sample} & \multicolumn{6}{c|}{Fitting Bias}                          \\
                            & \multicolumn{3}{c|}{SLD ($10^{-6} \text{\AA}^{-2}$)}
                            & \multicolumn{3}{c|}{Thickness ($\text{\AA}$)}              \\
                            & Layer 1 & Layer 2 & Layer 3 & Layer 1 & Layer 2 & Layer 3  \\ \hline
    Uncorrelated            & -0.003  & -0.002  & N/A     & -0.011  & 0.010   & N/A      \\ \hline
    Correlated              & -0.012  & -0.054  & 0.012   & -0.556  & 0.286   & 0.264    \\ \hline
    \end{tabular}
    \caption{Fitting biases in layer SLDs and thicknesses for the uncorrelated and correlated parameter samples of figure \ref{fig:confidence_ellipses}.}
    \label{tab:biases}
\end{table}

\subsection{Time Dependence}
One potential use of the FI in reflectometry is enabling early stopping of experiments based on counting statistics. To determine the feasibility of this application, and to validate our implementation, we investigated how parameter uncertainties change with measurement time. As derived in the SI, we should expect the uncertainty of a parameter, $\epsilon$, to be inversely proportional to the square root of the experiment measurement time, $\tau$. By using the fact that $\epsilon \propto 1/\sqrt{\tau}$, introducing a non-zero proportionality constant, $\alpha$, and taking the natural logarithm of both sides we see
$$\begin{aligned}
    \ln{\epsilon} &= \ln{\frac{\alpha}{\sqrt{\tau}}} \\
                  &= \ln{\alpha} - \ln{\sqrt{\tau}} \\
                  &= -\frac{1}{2}\ln{\tau} + \ln{\alpha}
\end{aligned}$$
Using this result, we should expect the gradient of the plot of log parameter uncertainty, $\ln{\epsilon}$, vs. log time, $\ln{\tau}$, to be $-\frac{1}{2}$. To confirm this is the case, we compared established fitting uncertainty measures and uncertainties derived from the FI with increasing time using our experiment simulation framework; the fitting uncertainties were calculated using differential evolution in \texttt{refnx} and the FI uncertainties using equation \ref{eqn:FI_uncertainty}.

\begin{figure}
    \includegraphics[width=\textwidth]{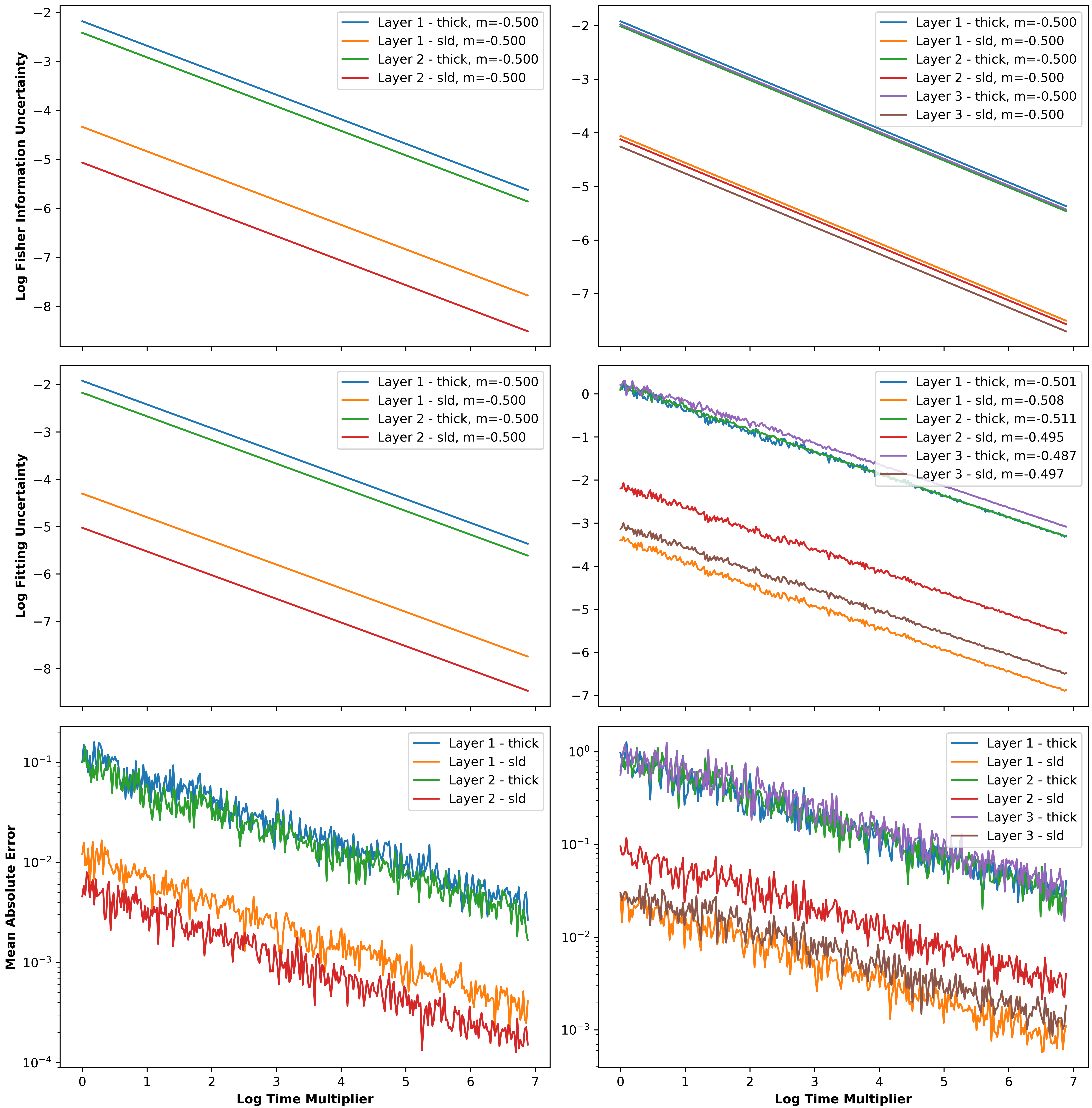}
    \caption{Log Fisher information uncertainty (top), log fitting uncertainty (middle) and mean absolute error (bottom) vs. log measurement time multiplier for each parameter of the mostly uncorrelated parameter sample (left) and correlated parameter sample (right); the uncertainties are taken as the mean from 10 simulated experiments for a given time multiplier. Included in the legends of the uncertainty time dependence plots are approximations of the gradients of the lines, $m$, as given by linear regression.}
    \label{fig:time_dependence}
\end{figure}

Using simple linear regression, we found that the time dependence for any parameter's uncertainty was indeed determined by the square root of the measurement time as is shown in figure \ref{fig:time_dependence}. We used the same samples and simulation parameters as those used for figure \ref{fig:confidence_ellipses} except for the simulated measurement time. Both samples were initially simulated using the same times as before and then these times were multiplied by an increasing ``time factor'' from 1 to 1000. This essentially split a fixed time-budget between the simulated angles, $0.7^{\circ}$ and $2.0^{\circ}$, with a ratio of 1:4.

For the simple sample, the relationship is perfectly exhibited. However, for the more complicated sample, the results are slightly noisier due to the increased difficulty of fitting. This is particularly noticeable when the counting time is low and the data being fitted is impacted by our added noise to a greater degree. With low counting statistics, differential evolution may terminate in a minimum of the $\chi^2$ parameter space that does not represent the ground truth model (I.e., the simulated data no longer uniquely describes the true parameter set), resulting in uncertainties that deviate from the time dependence relationship previously derived. This difficulty in fitting is shown in figure \ref{fig:time_dependence} where the mean absolute error between the ``ground truth'' and fitted parameter values are plotted against time. As can be seen at lower counting times, the fitting errors are approximately an order of magnitude larger than those at higher counting times.

Since we now know that parameter uncertainties decrease as the square root of the measurement time, we are able to easily project the evolution of these uncertainties and can predict when some desired threshold would be reached, at which time we may want to cease the measurement. It is the choice of the experimenter to decide on such a threshold with the choice likely weighing factors including time to change angle, time to change sample, total time budget and number of samples being measured. Such choices are not necessarily easy to make prior to starting an experiment and so automating this process may warrant further investigation.

\subsection{Application to Soft Matter} \label{softmatter_results}
\begin{figure}
    \includegraphics[width=\textwidth]{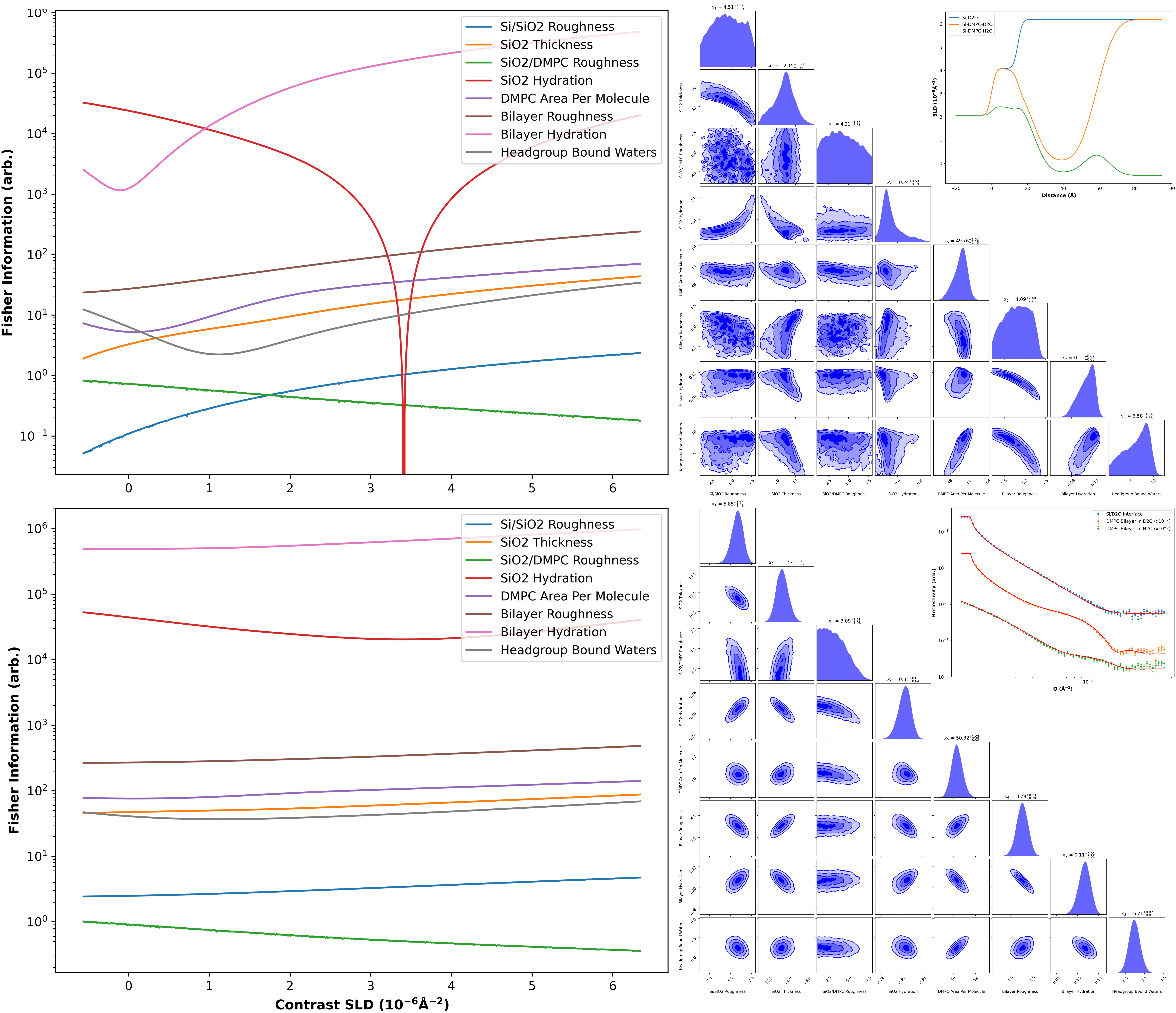}
    \caption{Fisher information vs. bulk water contrast SLD, for each model parameter of the reparameteristed DMPC bilayer model, for an initial contrast choice (top left) and second contrast choice (bottom left), assuming \ce{D2O} was measured first. Also shown are the nested sampling corner plots from sampling simulated data of solely \ce{D2O} (top right), and \ce{D2O} and \ce{H2O} contrasts (bottom right). Inset are the experimentally-fitted SLD profiles (top) and reflectivity curves (bottom), for the \ce{Si}/\ce{D2O} interface and the two solution isotopic contrast datasets, \ce{H2O} and \ce{D2O}, offset by factors of $10^{-2}$ and $10^{-3}$ respectively for clarity.}
    \label{fig:bilayer}
\end{figure}

As detailed in section \ref{softmatter_methods}, we applied our framework to a soft matter experiment by taking experimentally measured data, fitting a DMPC bilayer model, and reparameteristing the model as a function of bulk water contrast SLD. The fitted SLD profiles and experimental reflectivity data are shown in figure \ref{fig:bilayer}. Using the model, data was simulated for each contrast SLD from $-0.56 \times 10^{-6} \text{\AA}^{-2}$ to $6.35 \times 10^{-6} \text{\AA}^{-2}$ (pure \ce{H2O} to pure \ce{D2O}) and the FI calculated for each model parameter, obtained from the diagonal elements of the FI matrix of equation \ref{eqn:FI}. These results are shown in figure \ref{fig:bilayer} for an initial contrast choice and for a second contrast choice, assuming D2O was first measured. For simulation, angles of $0.7^{\circ}$ and $2.0^{\circ}$ and times of 15 and 60 minutes, respectively, were used (typical measurement times). Shown also are the nested sampling corner plots from sampling simulated data of solely \ce{D2O}, and \ce{D2O} and \ce{H2O} contrasts using the reparameterised model. Included in these corner plots are the sampling uncertainties associated with each parameter.

Since the units of the FI are \textit{nats} per parameter unit squared, it is not technically correct to directly compare the FI between parameters. However, we can still compare the information content of parameters of the same unit. As might naively be expected, we show that it is possible to extract more information about some parameters than others. This result is certainly no surprise for researchers in the field who have experience fitting this system, but it does allow us to quantify it.

We show that the information of a parameter as a function of contrast is non-monotonic, almost certainly due to hydration of various components leading to them becoming indistinguishable from neighbouring components for some bulk water deuterium concentrations. This is particularly noticeable with the \ce{SiO2} hydration parameter for an initial contrast choice, where the large drop in information is due to the matching of contrasts. Since most of the other model parameters describe multiple interfaces, with only one interface being able to become ``invisible'' through contrast matching at a time, there are no zeroes in the FI for these plots.

Figure \ref{fig:bilayer} indicates that the greatest information is almost always obtainable from the highest SLD water, \ce{D2O}. For an initial measurement contrast, the difference in information between the \ce{H2O} and \ce{D2O} extremes is significant. However, when considering a second measurement contrast, the information gain between the two contrast SLDs is less. It is well established that measuring multiple contrasts of different SLDs will reduce parameter uncertainties and so this result may seem unusual; it could suggest measuring \ce{D2O} for twice as long would reduce parameter uncertainties similarity to that of measuring \ce{D2O} and \ce{H2O}. However, we believe the aim of measuring multiple contrasts is not only in lowering parameter uncertainties but also in the decoupling of parameter correlations (I.e., lowering inter-parameter covariances). Therefore, only considering which contrasts maximise the FI will not account for this covariance reduction. This is illustrated in the corner plots of figure \ref{fig:bilayer} where the estimated posterior distributions from the \ce{D2O} and \ce{H2O} data are clearly much better defined (I.e., more Gaussian) than just \ce{D2O}. For example, the model roughness parameters are very poorly defined with just a single \ce{D2O} measurement.

While difficult to display due to the number of parameter pairs, if you have a model, it would be possible to calculate the optimal contrast to measure that minimises both parameter variances and inter-parameter covariances. The optimal solution is almost certainly model dependent, but given the broad features found here, it is unlikely to be an issue having a slightly incorrect model.

\subsection{Limitations}
Our framework is essentially frequentist and, much like one does in hypothesis testing, it proceeds by calculating probabilities based on hypothetical, assumed or estimated parameter values. The size of uncertainties, for example, are those that would exist, if the estimated parameter values were correct. Determining uncertainties in this way may appear to be an issue, particularly in reflectometry, since the determination of the globally ``correct'' model is non-trivial. As ever, it is the choice of the experimenter to decide whether their model, guided by their underlying knowledge of the system, will accurately represent the true system. However, even if the model is not exactly correct, the FI still provides value. Since our calculations effectively perform a sensitivity analysis of the parameters (more sensitive means a larger gradient in the Jacobian, $\mathbf{J}$, and therefore a larger FI), similar models with differing values are still very likely to have the same behaviour and give the same trends. For example, for our DMPC bilayer data, there are many more model parameterisations that have been argued in the literature, and differing values for the parameters we have chosen to fix. However, our simplifying assumptions describe the measured data to a satisfactory level and we are confident that equivalent parameterisations would not change the trends found from our calculations.

Usually, to optimise the parameter values, the model is fitted before calculating the FI; in our case, we have used differential evolution. Fitting algorithms can often provide estimates of parameter variances and covariances and in some cases these values may be very similar to those provided by our framework. However, the fit-derived (co)variances have no deeper statistical underpinning, and do not allow you to further investigate the system. Our framework, being based on the counting statistics and parameter sensitivity of the model, enables almost instant calculation of uncertainties measured at any point in the future, or for different contrasts/conditions. Determining the (co)variances from a fit several hundreds of times to create a ``phase diagram'', similar to those described in section \ref{softmatter_results}, is not feasible in anything close to real time, as would be useful during an experiment.

\section{Future Work}
The presented framework has many potential applications in neutron reflectometry, and other scattering techniques based on counting statistics. As demonstrated by our soft matter application, experimental design is one such use where the FI could be used to influence real-time decisions of measurement angle and/or contrast choice; a similar Bayesian approach would be infeasible for real-time application due to computational overhead. We would also like to apply our framework to more complex real-world systems, such as magnetic structures. This could provide answers to common questions posed in the literature and give insight as to why particular experimental design choices have found popularity.

The FI framework has the possibility of being extended to quantify additional factors of an experiment such as time to change sample or angle. Additionally, work on quantifying and incorporating fitting biases and inter-parameter correlations into the FI calculation and subsequent analysis could bridge the gap between our framework and established methods. Since the FI uncertainties do not always match those obtained from established methods, it could also be possible to provide experimenters with a metric detailing how closely the FI results would be expected match to established methods. Since the largest variations were found to occur when parameters are strongly correlated, the Pearson correlation coefficient applied to the FI matrix (or similar) could be indicative here.

\section{Conclusions}
In this work, we presented a framework for determining the maximum information gain and experimental design optimisation of neutron reflectometry experiments using the Fisher information (FI). We demonstrated how the FI allows us to quantify the information content of a measured data point in relation to given model parameters. To illustrate this point, we developed a robust framework for simulating experimental data with realistic noise characteristics, and then compared the FI-derived results to Bayesian sampling methods. The FI describes the maximum possible extractable information, and therefore can be significantly different from sampling methods. However, this approach has significant upsides in its run time, and ability to project uncertainties, as well as the ability to run experiments \textit{in silico}. Finally, we demonstrated a practical application of the approach in determining the information content of the parameters of a DMPC bilayer sample parameterised as a function of the bulk water contrast, allowing us to ascertain optimal measurement conditions.

The code for this work is open source and freely available on GitHub \cite{GitHub}.

\section{Acknowledgements}
This work has been partially supported by the STFC Facilities Programme Fund through the ISIS Neutron and Muon Source, and Scientific Computing Department of Rutherford Appleton Laboratory, Science and Technology Facilities Council, and by the Wave 1 of The UKRI Strategic Priorities Fund under the EPSRC Grant EP/T001569/1, particularly the ``AI for Science'' theme within that grant and The Alan Turing Institute. We would also like to thank Luke Clifton for his assistance and expertise in fitting the DMPC data.

\bibliographystyle{apalike}
\bibliography{main}

\newpage
\section{Supplementary Information}
\subsection{Deriving the Fisher Information}
\subsubsection{Notation}
\begin{center}
\begin{tabular}{rl}
$N$ & Number of bins or equivalently, data points \\
$M$ & Number of model parameters \\
$\lambda_i$   & Expected neutron counts in bin (parameter) $i$ \\
$\lambda = (\lambda_1 ... \lambda_N)$ & Vector of expected neutron counts in all bins (parameter)\\
$\tau_\theta$ & Time recording at experimental condition $\theta$, e.g. angle, contrast\\
$\mu_{i\theta}$ & Incident flux in bin $i$ for condition $\theta$ (parameter) \\
$g^{z}$ & Fisher information for parameterisation $z$ \\
$\xi_j$ & Parameters of the structure model \\
$r_i(\xi)$ & Reflectivity for bin $i$ from model with parameters $\xi$ \\
$X$ & Random variable describing a full measurement of $N$ points\\
$x_i \in \mathcal{X}_i$ & Neutron count in bin $i$ \\
$p(x; z)$ & Probability distribution for measurements $x$, parameterised by $z$ \\
$\Pr(\text{[event]})$ & Probability of event \\
$s_i$ & Total number of incident neutrons in bin $i$ \\
\end{tabular}
\end{center}

\subsubsection{The Model}
In reflectometry, a model describes the reflectivity at a given neutron momentum transfer. These momentum transfer values are binned with the measured number of neutrons in bin $i$ being given by
\begin{equation}
\label{eqn:lambdasigma}
    \lambda_i = r_i s_i(\tau)
\end{equation}
where $s_i$ is the number of neutrons incident in bin $i$. $s_i$ is a function of the number of incident neutrons in the $i$th bin at each experimental condition, $\mu_{i\theta}$, and the time each condition is measured for $\tau_\theta$.
$$s_i(\tau) = \sum_k \tau_k \mu_{ik}$$
and so
$$\lambda_i(\xi) = \sum_\theta r_i(\xi) \tau_\theta \mu_{i\theta}$$

\subsubsection{Fisher Information about the $\lambda$ coordinates}
The probability distribution of the measurement in one bin is Poisson distributed
$$\Pr(X_i = x_i; \lambda_i) = \frac{e^{-\lambda_i}\lambda_i^{x_i}}{x_i!}$$
and the corresponding Fisher information (FI) for this bin, with respect to $\lambda_i$, is
$$g^{\lambda_i} = 1/\lambda_i = 1/\expect[\var{X_i}]$$

The probability distribution for the whole measurement is, by independence
$$\Pr(X = (x_0 ... x_N)) = \prod_i Pr(X_i = x_i)$$
and the corresponding FI with respect to $\lambda$ is
$$g^\lambda_{jk} = \left\{\begin{array}{ll} g^{\lambda_k} & \text{if } j=k \\ 0 & \text{otherwise} \end{array} \right. $$
i.e. a diagonal matrix with values of $g^{\lambda_i}$ which happens to equal (in this case, but not in general), $\expect[\cov{X}]^{-1}$.

\subsubsection{Fisher Information about the $\xi$ coordinates.}
In general, we can transform the FI using tensor transforms, i.e.
$$ g^Z_{ij} = g^Y_{ab} \pdf[y_a]{z_i} \pdf[y_b]{z_j} $$
So, the FI in terms of $\xi$ is just
$$g^{\xi}_{ij} = g^\lambda_{ab} \pdf[\lambda_a]{\xi_i} \pdf[\lambda_b]{\xi_j}$$

To get the FI as a function of $\tau$, we take the derivative of equation \ref{eqn:lambdasigma} with respect to $\xi$ which gives us
\begin{equation}
\label{eqn:dlambda}
    \pdf[\lambda_i]{\xi_j} = s_i(\tau) \pdf[r_i]{\xi_j}
\end{equation}

The derivative $\pdf[r_i]{\xi_j}$ is obtained from the model alone (irrespective of the data); it is the derivative of the reflectivity for bin $i$ with respect to the $j$th $\xi$ parameter.

We can now put everything together. First, we have the initial FI about the $\lambda$ parameter, which we can write in terms of $s_i$ and $r_i$
$$\mathbf{g}^\lambda = \begin{bmatrix}
1/s_1 r_1 & 0           & \cdots       & 0  \\
0          & 1/s_2 r_2  &   \cdots     & 0  \\
\vdots          & \vdots           & \ddots & \vdots  \\
0          &0            & \cdots       &  1/s_{N} r_N \\
\end{bmatrix}$$

Equation \ref{eqn:dlambda} can be re-written in matrix form in terms of a diagonal matrix, $\mathbf{S}$, and the Jacobian matrix, $\mathbf{J}$, for the $N$ modelled reflectivity points, with respect to the $M$ parameters, $\xi$.
$$ \mathbf{S} = \begin{bmatrix}
s_1 & 0           & \cdots       & 0  \\
0          & s_2  &   \cdots     & 0  \\
\vdots          & \vdots           & \ddots & \vdots  \\
0          &0            & \cdots       &  s_{N}  \\
\end{bmatrix}$$

$$
\mathbf{J} = \begin{bmatrix}
\pdf[r_1]{\xi_1} & \pdf[r_1]{\xi_2} & \cdots & \pdf[r_1]{\xi_M}   \\
\pdf[r_2]{\xi_1} & \pdf[r_2]{\xi_2} & \cdots & \pdf[r_2]{\xi_M}  \\
\vdots           & \vdots           & \ddots & \vdots  \\
\pdf[r_N]{\xi_1} &\pdf[r_N]{\xi_2}  & \cdots & \pdf[r_N]{\xi_M}  \\
\end{bmatrix}
$$

The tensor transformation of $\mathbf{g}^\lambda$ in this notation is then
$$ \mathbf{g}^\xi = (\mathbf{SJ})^T \mathbf{g}^\lambda (\mathbf{SJ}) = \mathbf{J}^T \mathbf{Sg}^\lambda \mathbf{SJ}$$
$$(M \times M ) = (M \times N ) (N \times N ) (N \times N ) (N \times N ) (N \times M )$$
and the matrix $\mathbf{Sg}^\lambda\mathbf{S}$ is a composition of diagonal matrices, and is equal to
$$\begin{bmatrix}
s_1 / r_1 & 0           & \cdots       & 0  \\
0          & s_2 / r_2  &   \cdots     & 0  \\
\vdots          & \vdots           & \ddots & \vdots  \\
0          &0            & \cdots       &  s_{N} / r_N \\
\end{bmatrix}$$

\subsubsection{Summary}
In summary, the FI about $\xi$ is given by
$$\mathbf{g}^\xi = \mathbf{J^T}\mathbf{MJ}$$

where, $\mathbf{J}$ is the Jacobian of the reflectances, $r_i$, with respect to the parameters, $\xi$. $\mathbf{M}$ is a diagonal matrix with entries $(s_0/r_0, s_1/r_1 \ldots s_N/r_N)$ and $s_i$ is the incident neutron flux, which depends on the experimental condition $k$ and the time spent measuring it, $\tau_k$
$$s_i(\tau) = \sum_k \tau_k \mu_{ik}$$

\subsection{Confidence Ellipses}
To calculate confidence intervals in the general case, we can find the set of parameters that differ from the estimate by a certain number of standard deviations. To do this, we need to know the length of a vector in parameter space in terms of the number of standard deviations. This is what the FI does (technically, it gives a linear approximation to the informational distance between distributions, but they are related). So, if we want to find a vector with a given length, $k$, we solve
$$k^2 = \Delta\mathbf{\xi}^T \mathbf{g} \Delta\mathbf{\xi} $$

It so happens that $k$ in the above equation can be interpreted as ``number of standard deviations'', so a $2\sigma$ error bar will have $k=2$. In practice, it is useful to fix a direction, and calculate the magnitude of the vector needed to reach the threshold. I.e. let $\Delta\xi = \epsilon\widehat{\Delta\xi}$ where $\widehat{\text{hat}}$ denotes a unit vector. Consider the 1-D case; $k^2 = \epsilon^2 g$, $g$ is analogous to the inverse variance, so we have $k^2 = \epsilon^2 / \sigma^2$. Therefore, if we want to know where $\epsilon = 2\sigma$, we would have $k^2 = (2\sigma)^2/\sigma^2$ and thus, $k=2$.

In 2-D, the unit vectors can be written as $(\sin \vartheta, \cos\vartheta)$. We can graphically solve the following
$$
k^2 = \epsilon^2 \underbrace{
  \left(
      \begin{bmatrix}\sin\vartheta, \cos\vartheta\end{bmatrix}
      \mathbf{g}^\xi
      \begin{bmatrix}\sin\vartheta\\\cos\vartheta\end{bmatrix}
  \right)}_{\text{scalar}}
$$
for $\epsilon$ over a sample of angles, $\theta$, in $[0,2\pi]$ by plotting the points $(\epsilon(\theta)\sin\theta, \epsilon(\theta)\cos\theta)$. The result is a confidence ellipse of size $k$ between two chosen parameters. If $|\xi| > 2$, then for two chosen parameters, $\xi_i$ and $\xi_j$, the above equation becomes
$$
k^2 = \epsilon^2
  \left(
      \begin{bmatrix}\sin\vartheta, \cos\vartheta\end{bmatrix}
      \begin{bmatrix}
          \mathbf{g}^\xi_{i,i} & \mathbf{g}^\xi_{i,j} \\
          \mathbf{g}^\xi_{j,i} & \mathbf{g}^\xi_{j,j}
      \end{bmatrix}
      \begin{bmatrix}\sin\vartheta\\\cos\vartheta\end{bmatrix}
  \right)
$$

\subsection{Point Estimates vs Posterior Distributions}
An estimator, usually written with a $\widehat{\text{hat}}$, is a function of sampled data that provides an estimate of a parameter. In frequentist statistics, one is concerned with the probability distributions of \emph{estimators}, not of the parameters themselves (as would be the case in Bayesian statistics). The distribution of estimators often has a variance. When it does, and when it is unbiased, i.e.
$$\expect[\hat{\xi}] = \xi$$
then it is related to the FI by the Cram\'{e}r-Rao bound:
$$ \var{\hat{\xi}} \succcurlyeq \frac{1}{N} (\mathbf{g}^\xi)^{-1}$$

\subsection{Time Dependence}
If there is a single measurement condition, we have $s_i = \mu_i\tau$. Then, we see that $\tau$ is a factor of all the $s_i$'s and so of the matrix $\mathbf{M}$ and thus of $\mathbf{g}^\xi$. We can therefore write $\mathbf{g}^\xi = \tau \mathbf{f}^\xi$. The size of an ``error bar'', $\epsilon$, for a single parameter, $\xi$, at a threshold, $k$, in a direction, $\widehat{\Delta\xi}$, is given by
$$k^2 = (\epsilon \widehat{\Delta\xi}) \mathbf{g}^\xi (\epsilon \widehat{\Delta\xi})$$
And so,
$$k^2 = (\epsilon \widehat{\Delta\xi}) \tau \mathbf{f}^\xi (\epsilon \widehat{\Delta\xi})$$
Hence $\epsilon \propto 1/\sqrt{\tau}$.

\subsection{Bilayer Model Parameterisation}
For the measured DMPC sample, each experimental dataset was recorded with an instrument resolution of 2\%. The instrument backgrounds were $3.21 \times 10^{-6}$, $2.80 \times 10^{-6}$ and $2.06 \times 10^{-6}$ respectively. The SLDs of the \ce{Si} substrate and following \ce{SiO2} layer were defined using known values of $2.073 \times 10^{-6} \text{\AA}^{-2}$ and $3.41 \times 10^{-6} \text{\AA}^{-2}$. From fitting in \texttt{RasCAL}, we obtained values of $14.7 \text{\AA}$ and 24.5\% for the SiO2 layer's thickness and hydration. For both the \ce{Si}/\ce{SiO2} and \ce{SiO2}/DMPC interfacial roughnesses, we obtained $2.00 \text{\AA}$; all other interfacial roughnesses shared a common parameter that was fitted as $6.57 \text{\AA}$.

We have assumed the molecular volumes of the headgroups and tailgroups are known and constant at $320.9 \text{\AA}^{3}$ and $783.3 \text{\AA}^{3}$, and that any changes in molecule surface area are inversely proportional to the headgroup and tailgroup thicknesses. Therefore, we need only fit one parameter: the area per molecule (APM) at the surface. From the APM we can calculate the tailgroups thickness using the known volume.
\begin{equation}
    \label{eqn:thickness}
    \text{Thickness} = \frac{\text{Volume}}{\text{APM}}
\end{equation}
From fitting the APM parameter, we obtained a value of $49.9 \text{\AA}^{2}$. Also using the tailgroup volume, the SLD of the tailgroups can be calculated using the known tailgroup scattering length (SL) of $-3.08 \times 10^{-4} \text{\AA}$ and equation
\begin{equation}
    \label{eqn:SLD}
    \text{SLD}(\rho) = \frac{\Sigma b}{\text{Volume}}
\end{equation}

Both the headgroups and tailgroups contain water through defects across their surfaces but there is also water bound to the hydrophilic headgroups. The model accounted for these differing hydration types by varying two parameters: the total bilayer hydration and headgroup bound waters for which we obtained values of 7.37\% and 3.59. The headgroup water SLs in \ce{H2O} and \ce{D2O} were calculated as the product of the bound waters parameter and the known SLs of $-1.64 \times 10^{-5} \text{\AA}$ and $2.00 \times 10^{-4} \text{\AA}$ for the \ce{H2O} and \ce{D2O} solutions. Further, the product of the bound waters parameter and the known volume of water, $30.4 \text{\AA}^{3}$, yielded the headgroup water volume. The headgroup thickness was calculated using the total headgroup volume (including bound water), APM parameter and equation \ref{eqn:thickness}. To determine the headgroup SLs in \ce{H2O} and \ce{D2O} we took the sum of the known headgroup SL of $6.41 \times 10^{-4} \text{\AA}$ and headgroup SL in \ce{H2O} and \ce{D2O} previously calculated. Finally, using the headgroup SL in each contrast and the calculated headgroup volume, we calculated the SLD of the headgroup in each solution using equation \ref{eqn:SLD}.

We reparamaterised the fitted model as a function of bulk water contrast SLD using the known SLD of the DMPC headgroups of $1.98 \times 10^{-6} \text{\AA}^{-2}$ (if no hydrating water was present) and approximating the headgroup hydration at 27\%.

\end{document}